\documentclass[%
 reprint,
 amsmath,amssymb,
 aps,
]{revtex4-2}
\usepackage{braket}
\usepackage{dsfont}
\usepackage{graphicx}% Include figure files
\usepackage{dcolumn}% Align table columns on decimal point
\usepackage{bm}% bold math
\usepackage{subcaption}
\usepackage{cleveref}
\usepackage{xcolor}

\begin{document}

% \preprint{APS/123-QED}

\title{Stability of quantum many-body scars on PXP model}% Force line breaks with \\

\author{Alessandra Chioquetta}
\email{alessandrachioquetta@ufmg.br}
\affiliation{Departamento de Física, Universidade Federal de Minas Gerais, Belo Horizonte, MG, Brazil}

\author{Raphael C. Drumond}
\affiliation{Departamento de Matemática, Universidade Federal de Minas Gerais, Belo Horizonte, MG, Brazil}
% \affiliation{%
%  Authors' institution and/or address\\
%  This line break forced with \textbackslash\textbackslash
% }%

% \collaboration{MUSO Collaboration}%\noaffiliation

%  \homepage{http://www.Second.institution.edu/~Charlie.Author}
% \affiliation{
%  Second institution and/or address\\
%  This line break forced% with \\
% }%
% \affiliation{
%  Third institution, the second for Charlie Author
% \affiliation{%
%  Authors' institution and/or address\\
%  This line break forced with \textbackslash\textbackslash
% }%

% \collaboration{CLEO Collaboration}%\noaffiliation

\date{\today}% It is always \today, today,

\begin{abstract}
We investigate the stability of quantum many-body scars under perturbations, within the PXP model. We numerically compute the fidelity and average correlations to monitor the state evolution and to identify revivals. The results indicate that, on the one hand, the entanglement entropy of PXP scars exhibit great sensitivity, in the sense that their profile approaches the ones expected for thermal states already for very small perturbations. On the other hand, other scar signatures, such as the revivals of states having large overlap with scars, show remarkable robustness. Additionally, we examined the effects of minor disturbances on initial states that previously exhibited high overlap with scars and consistent revivals. Our analysis revealed that different types of disturbances can induce markedly different behaviors, such as partially ``freezing'' the chain, leading to sustained oscillations, or accelerating the process of thermalization.

% Secondary publications and information retrieval purposes.
% You may use the \texttt{description} environment to structure your abstract;
% use the optional argument of the \verb+\item+ command to give the category of each item. 
% \end{description}
\end{abstract}

\maketitle

%\tableofcontents

\section{\label{sec:level1}Introduction}
Statistical mechanics assumes an isolated system will eventually relax towards a state of thermal equilibrium and maximum allowed entropy. However, at a first glance, this concept conflicts with the idea that the von Neumann entropy of a many-body system in a pure state must remain zero under a unitary dynamics. The phenomenon of quantum thermalization explains how statistical mechanics emerges in such systems: local sub-regions reach a thermal state through the exchange of quantum correlations with the rest of the system. This process results in a pure state becoming locally indistinguishable from a thermal state, as described by the eigenstate thermalization hypothesis (ETH) \cite{deutsch1991quantum,srednicki1994chaos, rigol2012alternatives, langlett2022rainbow}.

Nevertheless, there are violations of the ETH, such as in quantum integrable systems. These systems avoid thermalization due to a large number of conservation laws, but they are susceptible to perturbations \cite{abanin2019colloquium,nandkishore2015many}. More resilient forms of ETH violations are found in strongly disordered interacting systems. These systems exhibit phenomena such as many-body localization. Recently, experiments with cold atoms have shown that some initial states in many-body models exhibit substantial revivals during the thermalization process \cite{bernien2017probing}. In these cases, the observables return to their initial values, leading to continued oscillations and a prolonged deviation from thermalization. It was then suggested~\cite{turner2018weak} that such phenomenon could be attributed to the existence of a few exceptional eigenstates in the many-body Hamiltonian, that had, moreover, a higher than usual overlap with the initial product state considered in the experiment. Such eigenstates were dubbed quantum many-body scars due to their conceptual relation with the quantum scared states that few-body chaotic quantum systems may exhibit~\cite{heller1984bound}.

The discovery of quantum many-body scars has sparked a wider interest in the study of ergodicity breaking across different types of quantum systems. This has led to a greater understanding of this complex phenomenon and may hold potential implications for a range of fields, from condensed matter physics to quantum information science \cite{gillmeister2020ultrafast, hu2014optically, yuan2022quantum,zhang2023many,wildeboer2021topological,mcclarty2020disorder}. 

Studies \cite{khemani2019signatures, lin2020slow} have addressed the stability and thermalization behavior of these unique states under perturbations. Particularly, these investigations center around identifying terms that drive the system towards an integrable regime, thereby revealing a parent Hamiltonian capable of governing the early-to-intermediate time dynamics of the experimental system.  On the other hand, actual systems will exhibit generic disturbances that will eventually guide the system towards thermalization.

Aiming to understand how scar behave under the disturbances of the environment, in this article we explore the impact of disordered disturbances applied to the PXP model, as well as errors in the initial state, specifically to address the robustness of many-body scars under generic perturbations.

Our numerical analysis focuses on fidelity and average correlations to monitor state evolution and identify revivals. The results reveal that while the entanglement entropy of PXP scars is highly sensitive to perturbations—approaching thermal state profiles even for small perturbations—the revivals of states with large overlaps with scars display remarkable robustness. Additionally, we find that minor disturbances to initial states with high scar overlap can lead to significantly different behaviors, such as ``freezing'' parts of the chain that sustain oscillations or, in the other case, accelerate the thermalization.

In our first analysis, where we include disordering terms into the Hamiltonian, we considered both period-2 $\ket{\mathds{Z}_2}$ and period-3 $\ket{\mathds{Z}_3}  = \ket{\bullet \circ \circ \bullet \circ \circ ...}$ density wave states,  which have been examined in prior studies of quantum scarring \cite{turner2018quantum, serbyn2021quantum} and can be effectively generated in experiment with Rydberg atoms~\cite{bernien2017probing}. We moreover included the state $\ket{0} = \ket{\circ \circ \circ ...}$ in the study for comparison, since it has no significant overlap with the scars of the PXP model and is expected to thermalize. In the second part, we investigated modifications on $\ket{{\mathds{Z}_2}}$ that leads to two significantly distinct behaviors.

The structure of this paper is as follows: In Sec. \ref{model}, we introduce the PXP model, describe the boundary conditions we considered, and discuss properties related to special states. Sec. \ref{disHam} details the disordering terms applied to the model, evaluates the robustness of three initial states under these conditions, and examines the behavior of decaying fidelity as the perturbation strength increases. The second approach is explored in Sec. \ref{init_defects}, where we introduce errors into the initial state with the most prominent revivals, resulting in two distinct configurations with different behaviors. We conclude with a discussion in Sec. \ref{conclusion}.

\section{\label{model}The PXP model}

The first many-body scars were reported in an experiment with Rydberg atoms \cite{bernien2017probing}, which, under the Rydberg blockade constrain, can be approximated by the so-called PXP model:
\begin{equation}
    H = \sum_{i=1}^{N} P_{i-1}X_{i}P_{i+1},
    \label{PXP}
\end{equation}
where $P_i = \ket{\circ}\bra{\circ}$ is the projector into the spin down state for site $i$ and $X_i = \ket{\circ}\bra{\bullet} + \ket{\bullet}\bra{\circ}$ is the Pauli operator on the $x$-direction. This Hamiltonian preserves the subspace where first neighbors are not simultaneously in the excited state. Therefore, initial states in that subspace space will remain in it. States with two neighboring atoms in the excited state, such as $\ket{\bullet \bullet \bullet}$, $\ket{\circ \bullet \bullet}$ and $\ket{\bullet \bullet \circ}$, are considered ``prohibited''. 

We investigated both open boundary conditions (OBC) and periodic boundary conditions (PBC). Specifically for OBC case, the boundary terms were included
 \begin{equation}
     H_{OBC} = X_1P_2 + P_{N-1}X_{N}.
 \end{equation}

 In our analysis and discussions, we consider both cases. However, the results were observed to be qualitatively the same, so we will present data mainly for PBC.

Experiments \cite{bernien2017probing} and numerical simulations on small systems \cite{sun2008numerical, olmos2009collective} have shown that the relaxation under unitary dynamics of this model strongly depends on the initial state. In particular, the period-2 density wave states
\begin{equation}
    \ket{\mathds{Z}_2} = \ket{\bullet \circ \bullet \circ ...}, \;\;\;\;\; \ket{\mathds{Z}_2^{\prime}} = \ket{\circ \bullet \circ \bullet...}
    \label{z2}
\end{equation}
 revealed long-time oscillations of local observables, such as the correlation between atoms, and relatively slower entanglement entropy growth. Similarly, although to a lesser extent, states such as $\ket{\mathds{Z}_3} = \ket{\bullet \circ \circ \bullet \circ \circ ...}$  displayed slower entanglement entropy growth as well \cite{turner2018weak, bernien2017probing}.

This behavior can be attributed to the presence of scar states in the PXP model, which  are eigenstates with approximately the same energy spacing and characterized by low entanglement entropy compared to typical eigenstates. This near-equidistant energy spacing results in constructive interference during the system's evolution, leading to periodic pronounced revivals of the initial state manifested through the fidelity. Specifically, the $\ket{\mathds{Z}_2}$ and $\ket{\mathds{Z}_3}$ states exhibit a high overlap with these scar states, resulting in their distinctive non-thermal behavior.

\section{\label{disHam}Disordered Hamiltonian}

The first disturbance we considered was introducing disordering terms into the Hamiltonian:
\begin{equation}
     H_{p} = H_{PXP} \; + \sum_{i=1}^{N} (h_{X}^i X_{i} + h_{Y}^i Y_{i} + h_{Z}^i Z_{i}) \;,
     \label{perturbH}
\end{equation}
where $X_{i}$, $Y_{i}$ and $Z_{i}$ are the Pauli matrices acting on site $i$, and $h_{\sigma}^{i}$ are i.i.d. random variables, uniformly distributed from $-W/2$ to $W/2$, with $W>0$ being the strength of disorder. The idea is to consider the simplest kind of disordered perturbation (single sited) that do not, moreover, preserve the constrained subspace of the model.

We investigate mainly chains with $N=18$ sites,  a choice that allowed us to numerically evolve the full many-body quantum state~\footnote{We use QuTip \cite{JOHANSSON20121760} solver to evolve the initial states.}. We considered perturbation strengths ranging from $W=0$ to $W=0.5$ in the Hamiltonian Eq.~\eqref{perturbH}.  
Note that are no conserved subspaces under the perturbations considered. Moreover, the number of sites is particularly suitable, since it meets the necessary conditions of an even number of sites for $\ket{\mathds{Z}_2}$ and a multiple of three for $\ket{\mathds{Z}_3}$ states under PBC's symmetry.
For further investigation, however, we employed different chain sizes, especially in cases where we needed to exactly diagonalize the full Hamiltonian.
To evaluate the robustness of the states, we computed the fidelity,
\begin{equation}
    F(t) = |\bra{\psi(0)}\psi(t)\rangle|^2,
\end{equation}
 checking how much the evolved state overlaps with the initial one over time, indicating the existence of revivals. We moreover computed the spatial average correlation
\begin{equation}
    \langle Z_{i} Z_{i+1} \rangle = \frac{1}{N} \sum_{i=1}^{N-1} Z_{i} Z_{i+1},
\end{equation}
to verify observable oscillations, which is another indication of the existence of scars.

\begin{figure}[ht!]
    \centering
    \begin{subfigure}[b]{0.5\textwidth}
        \centering
        \includegraphics[scale=0.22]{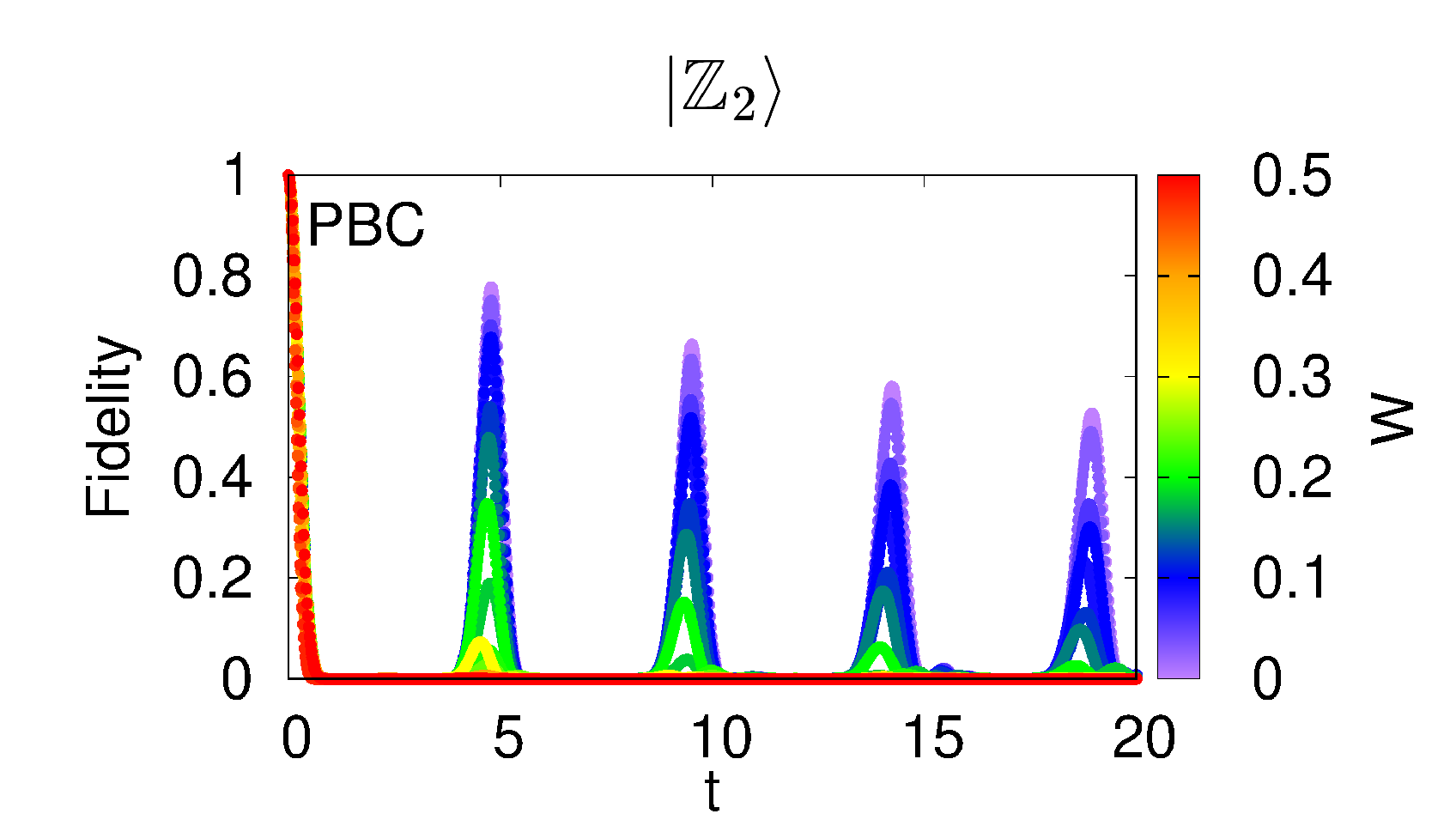}
        \caption{}
        \label{OverlapZ2pbc}
    \end{subfigure}%
    \hfill 
    \begin{subfigure}[b]{0.5\textwidth}
        \centering
        \includegraphics[scale=0.22]{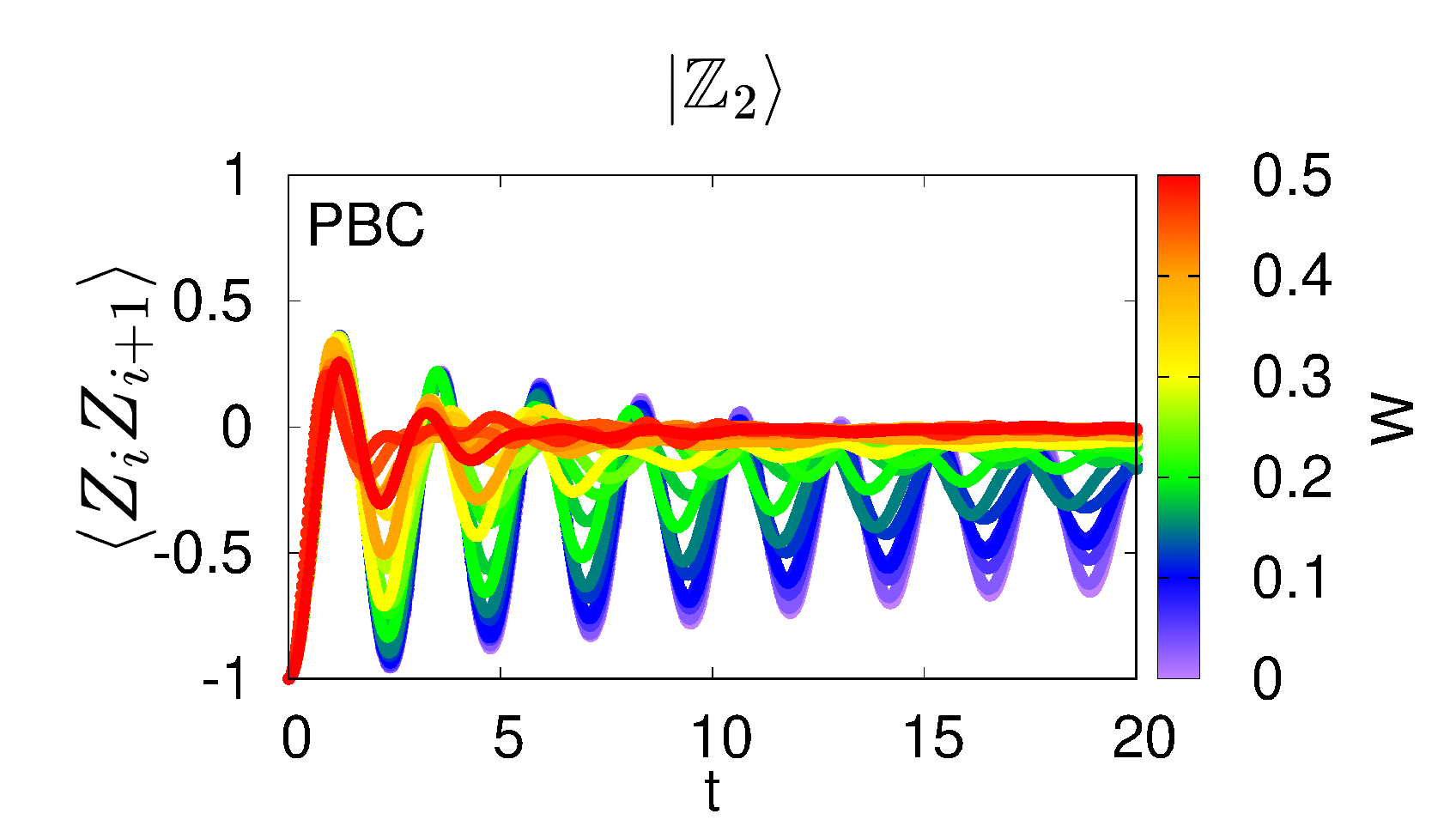}
        \caption{}
        \label{corrZ2pbc}
    \end{subfigure}
    \caption{(a) Fidelity and (b) average correlation for initial state $\ket{\mathds{Z}_2}$ in a chain of 18 sites with periodic boundary conditions.}
    \label{z2pert}
\end{figure}

As expected, with increasing strength of the perturbation, the initial state $\ket{\mathds{Z}_2}$ demonstrated diminishing revivals over time and with increasing perturbation strength. For perturbations with $W<0.1$, the fidelity retained significant revivals, indicating that the system's state maintains a high overlap with the initial state during its evolution. This suggests that the quantum many-body scars are relatively robust to small perturbations, as evidenced in \Cref{OverlapZ2pbc}.
In addition to fidelity, the average correlation measurements further support this behavior. Clear long-term oscillations were observed within the range $0\leq W\lesssim 0.1$ (\Cref{corrZ2pbc}).

\begin{figure}[hbt!]
    \centering
    \begin{subfigure}[b]{0.5\textwidth}
        \centering
        \includegraphics[scale=0.22]{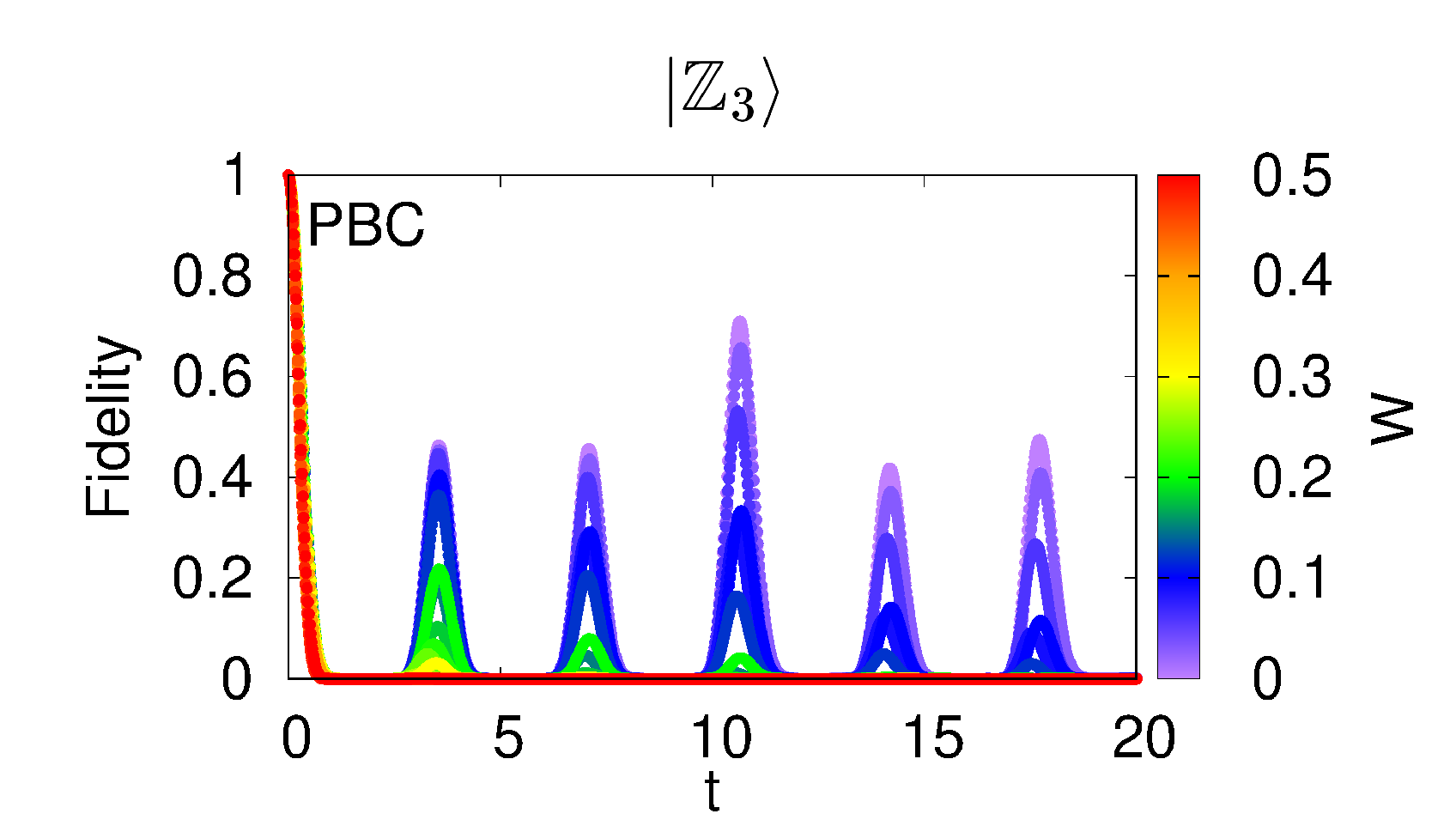}
        \caption{}
        \label{OverlapZ3pbc}
    \end{subfigure}%
    \hfill 
    \begin{subfigure}[b]{0.5\textwidth}
        \centering
        \includegraphics[scale=0.22]{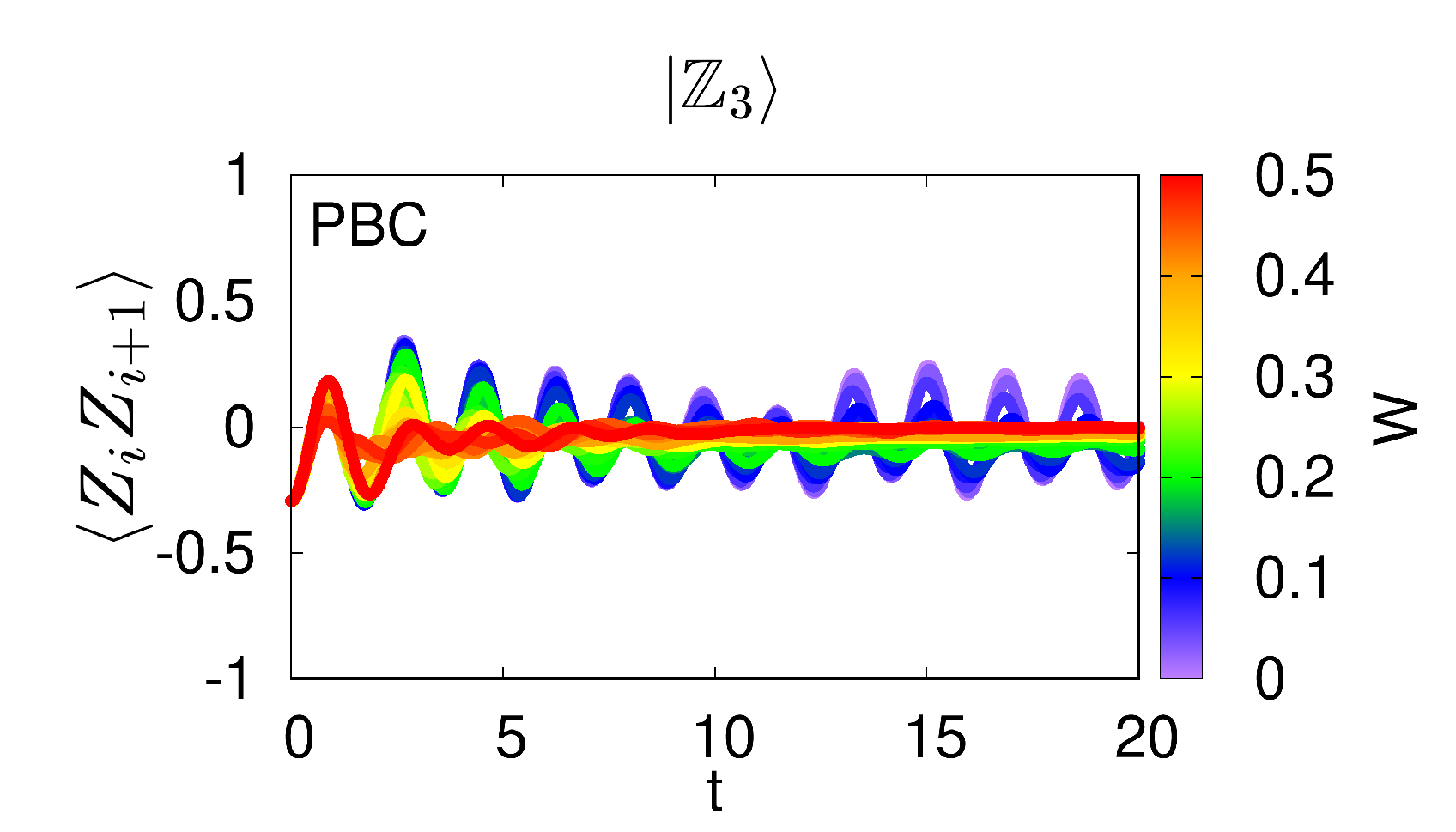}
        \caption{}
        \label{corrZ3pbc}
    \end{subfigure}
    \caption{(a) Fidelity and (b) average correlation for initial state $\ket{\mathds{Z}_3}$ in a chain of 18 sites with periodic boundary conditions.}
    \label{z3pert}
\end{figure}

The $\ket{\mathds{Z}_3}$ state under perturbations shows distinct behavior from   $\ket{\mathds{Z}_2}$. Specifically, for small values of the perturbation strength, $\ket{\mathds{Z}_3}$ exhibits pronounced revivals in fidelity at later times.  This delayed revival pattern suggests a different mechanism of coherence compared to $\ket{\mathds{Z}_2}$. However, as the perturbation strength increases, the fidelity of $\ket{\mathds{Z}_3}$ gradually decreases, showing a continuous reduction in revivals over time (\Cref{OverlapZ3pbc}), analog to what was observed in $\ket{\mathds{Z}_2}$ .
As for the average correlation, the measurements align with this behavior. Initially, the oscillations diminish, but they eventually restore their amplitude around the time when the most significant fidelity peak occurs (\Cref{corrZ3pbc}).

For perturbation strengths larger than $W > 0.1$, the revivals and oscillations, while still present, became less pronounced in both initial states considered. The system started to show more typical thermal behavior, stabilizing over shorter timescales. This transition suggests that with increasing perturbation strength, the system progressively aligns with the ETH, thereby losing the non-thermal features associated with many-body scars.

\begin{figure}[hbt!]
    \centering
    \begin{subfigure}[b]{0.5\textwidth}
        \centering
        \includegraphics[scale=0.22]{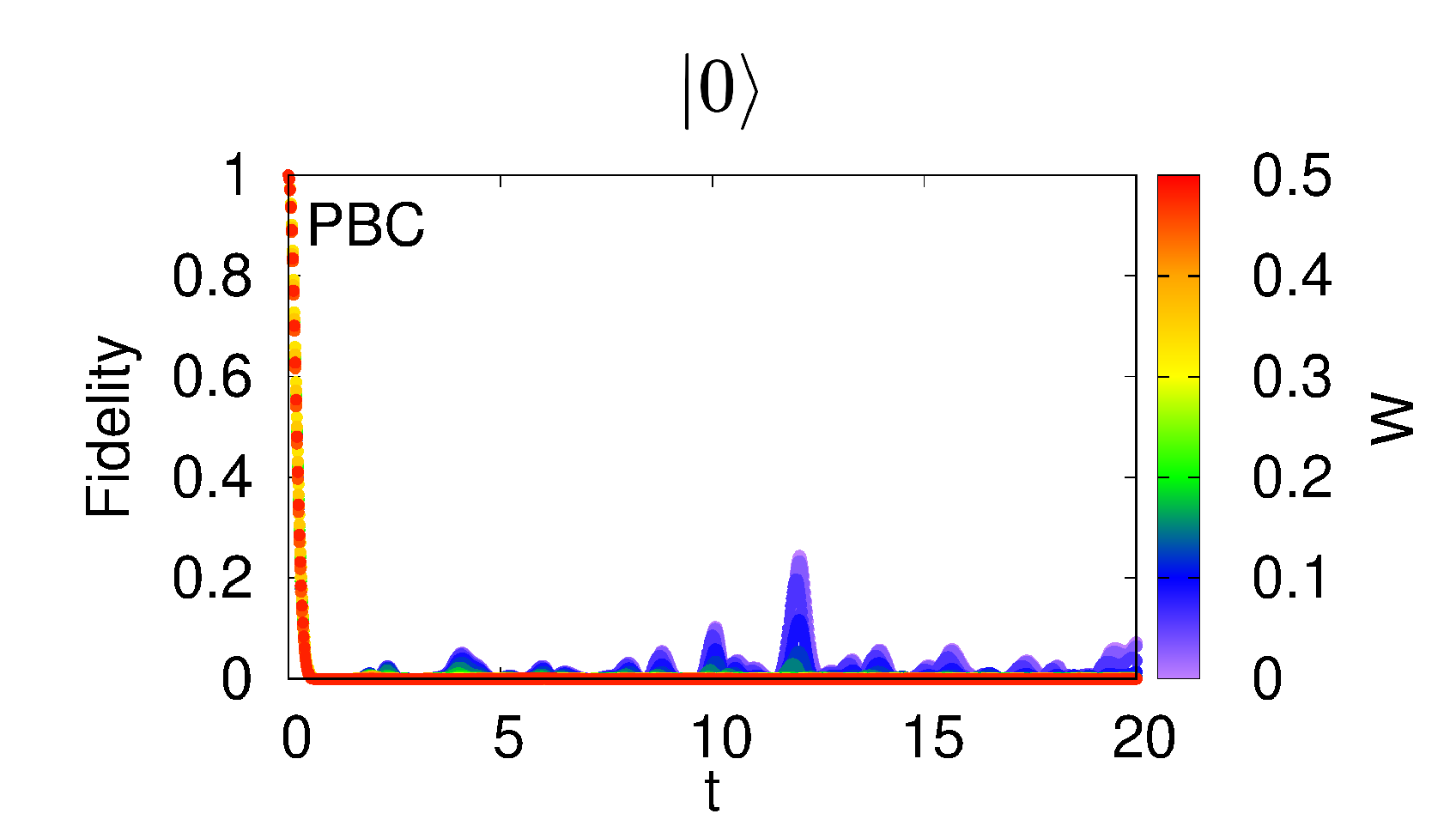}
        \caption{}
        \label{OverlapZ0pbc}
    \end{subfigure}%
    \hfill 
    \begin{subfigure}[b]{0.5\textwidth}
        \centering
        \includegraphics[scale=0.22]{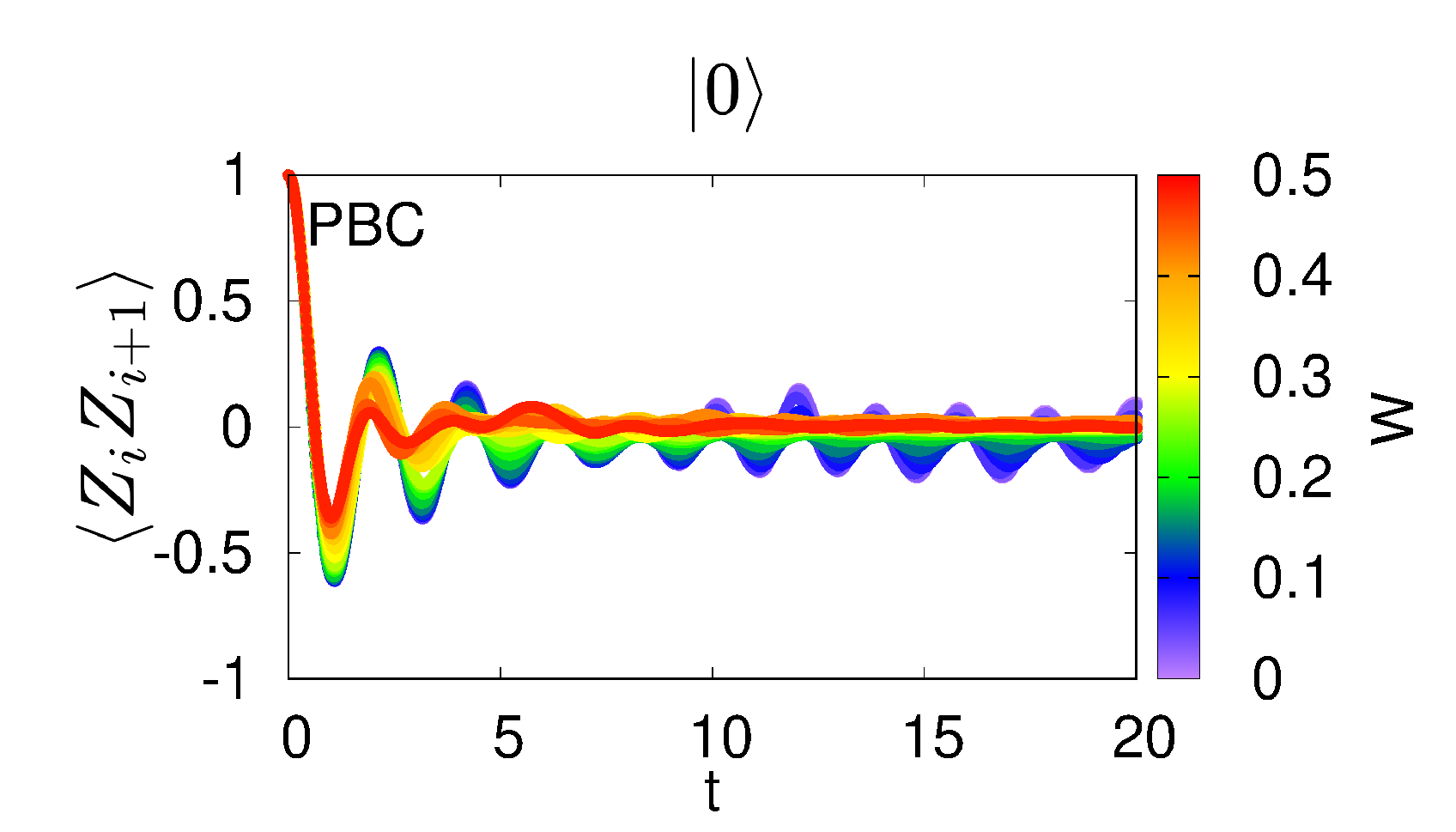}
        \caption{}
        \label{corrZ0pbc}
    \end{subfigure}
    \caption{(a) Fidelity and (b) average correlation for initial state $\ket{0}$ in a chain of 18 sites with periodic boundary conditions.}
    \label{z0pert}
\end{figure}

For comparison, \Cref{z0pert} illustrate the fidelity and average correlation of the state $\ket{0}$.This state displayed just small overlaps without any consistent revivals in fidelity over time. Regarding the correlations, there were minor oscillations with low values of $W$, but they quickly stabilized as the perturbation grows. As mentioned before, such state is expected to thermalize. Indeed, the entanglement entropy of $\ket{0}$ grows faster than $\ket{\mathds{Z}_2}$ and $\ket{\mathds{Z}_3}$, and it does not show significant overlap with PXP scars~\cite{turner2018weak}.

\begin{figure}[hbt!]
    \centering
    \includegraphics[scale=0.22]{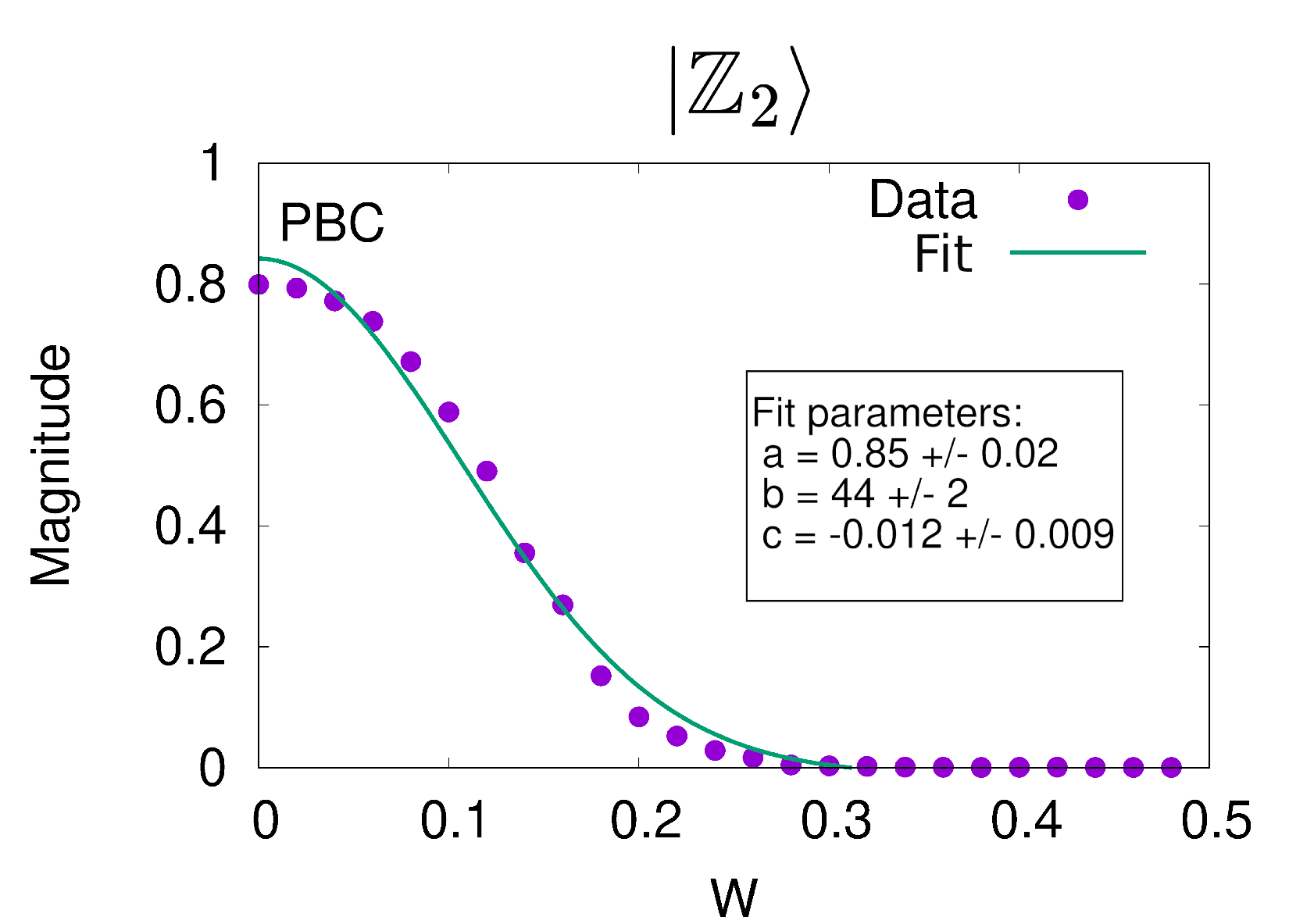}
    \caption{Magnitude of the most significant fidelity peak for state $\ket{\mathds{Z}_2}$, fitted with the function $a e^{-bW^2}+c$.}
    \label{magnitudeZ2}
\end{figure}

To investigate how fidelity depends on perturbation, we focused on the magnitude of the most significant fidelity peaks. Given that the values $h_{\sigma}^{i}$ are selected randomly, we averaged the dynamics over 10 realizations of disorder on chains with 16 sites for $\ket{\mathds{Z}_2}$.
When a quench is performed in a system, and it is allowed to thermalize, the fidelity decay over time typically follows a Gaussian profile in time and perturbation strength, $F(t)\approx e^{-c(Wt)^2}$, with $c$ being a constant~\cite{quan2006decay}. In the presence of scars, the time dependency clearly deviates from that approximation. It is interesting, however, that as a function of perturbation strength, the fidelity follows the same dependency, which is just a manifestation of the orthogonality catastrophe~\cite{anderson1967infrared}, as we can see in \Cref{magnitudeZ2}.

Additionally, we examined a 12-site chain with the initial state $\ket{\mathds{Z}_3}$, as a 16-site chain does not match the $\mathds{Z}_3$ symmetry under PBC. We found that its most significant peak exhibited similar behavior.

Under OBC, the fidelity and average correlation of $\ket{\mathds{Z}_2}$, the 18-site chain was similar to that observed with PBC, though with more rapidly diminishing magnitudes. In particular, the fidelity of the initial state $\ket{\mathds{Z}_3}$ demonstrated a consistent decrease over time, without the subsequent revivals characteristic of PBC. The average correlation followed this trend, with its oscillations stabilizing over time.

We also considered the entanglement entropy for the eigenstates of the perturbed Hamiltonians. However, the additional terms we included disrupt the constrained subspace, allowing states to disperse into the rest of the Hilbert space. In order to analyze the entropy in the subspace, we expressed the perturbed Hamiltonian in the basis of the constrained subspace, effectively incorporating the projectors $P_i$ into the perturbations:
\begin{equation}
    H_{p} = H_{PXP} \; + \sum_{i=1}^{N} P_{i-1}(h_{X}^i X_{i} + h_{Y}^i Y_{i} + h_{Z}^i Z_{i})P_{i+1}.
\end{equation}

\begin{figure}[hbt!]
    \centering
    \includegraphics[scale=0.5]{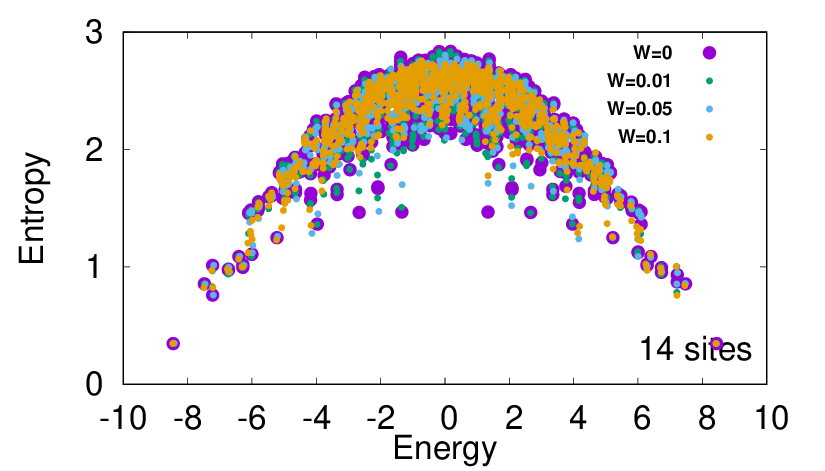}
    \caption{Entanglement entropy on the subspace of constrained Hamiltonians, with perturbation strength growing up to $W = 0.1$ ($W=0$ is the unperturbed PXP).}
    \label{entropyPert}
\end{figure}

This approach allows us to observe changes in the structure of scars as the perturbation strength $W$ varies, while preserving the scars within the constrained subspace. As illustrated in \Cref{entropyPert}, the distribution of entanglement entropy for a 14-site chain shifts as $W$ increases. For $W = 0$, the entanglement entropy values are lower, indicating the presence of many-body scar states that deviate from typical thermal states. However, as $W$ increases, the entropy distribution broadens and moves closer to that of thermal states. This progression demonstrates how increasing perturbation strength $W$ induces greater thermalization within the system, thereby diminishing the distinct scar structure.

%%%%%%%%%%%%%%%%%%%%%%%%%%%%%%%%%%%%%%%%%%%%%%%%%%%%%%%%%%%%%%%%%%%%
\section{Initial states with defects \label{init_defects}}

It is important to recall that, even in fully ergodic systems, thermalization is expected to take place only for a set of physically sensible states, namely, those with finite energy density, but small energy fluctuation~\cite{d2016quantum}. Indeed, a coherent superposition of two eigenstates, with macroscopically distinct energies, would never thermalize. However, states that do thermalize are, nevertheless, stable under local perturbations. If a state thermalizes, flipping a single spin of it will not change its energy density or overall energy fluctuation, so it will also thermalize. It is not clear, however, what to expect for non-thermal states.  

Here, we investigate how local perturbations in the initial state could affect the system's overall dynamics. We consider ``defects'' in the initial state $\ket{\mathds{Z}_2}$, rather than perturbing the Hamiltonian. Specifically, we flip one spin around the middle of the chain, resulting in two possible configurations:
\begin{equation*}
    \ket{\mathds{Z}_2^{'}} = \ket{...\; \bullet \circ  \; \bullet \underline{\bullet} \; \bullet \circ  \;...},
\end{equation*}
\vspace{-0.5cm}
\begin{equation*}
    \ket{\mathds{Z}_2^{''}} = \ket{...\; \bullet \circ  \;  \underline{\circ} \circ \;\bullet \circ  \; ...}.
\end{equation*}

We chose to study defects in the $\ket{\mathds{Z}_2}$ chain because it provides richer exploration opportunities for two main reasons. First, the $\ket{\mathds{Z}_2}$ configuration fits perfectly on any chain with an even number of sites, which simplifies symmetry checks under PBC. Second, the periodicity of the $\ket{\mathds{Z}_2}$ pattern is shorter and repeats more frequently within the chain compared to the $\ket{\mathds{Z}_3}$ pattern. This increased repetition in smaller chains makes the $\ket{\mathds{Z}_2}$ chain more suitable for detailed studies of defect impacts, as it allows for more consistent and observable periodic behavior.

In the first scenario, where we flipped a spin from down to up, the resulting state falls outside the constrained subspace. This flip ``freezes'' part of the chain, not only stabilizing the three consecutive up states, but also affecting their immediate down neighbors, as they cannot flip up. This stabilization creates a localized region of  five fixed spins, altering the dynamics of the chain significantly.

\begin{figure}[hbt!]
    \centering
    \begin{subfigure}[b]{0.5\textwidth}
        \centering
        \includegraphics[scale=0.21]{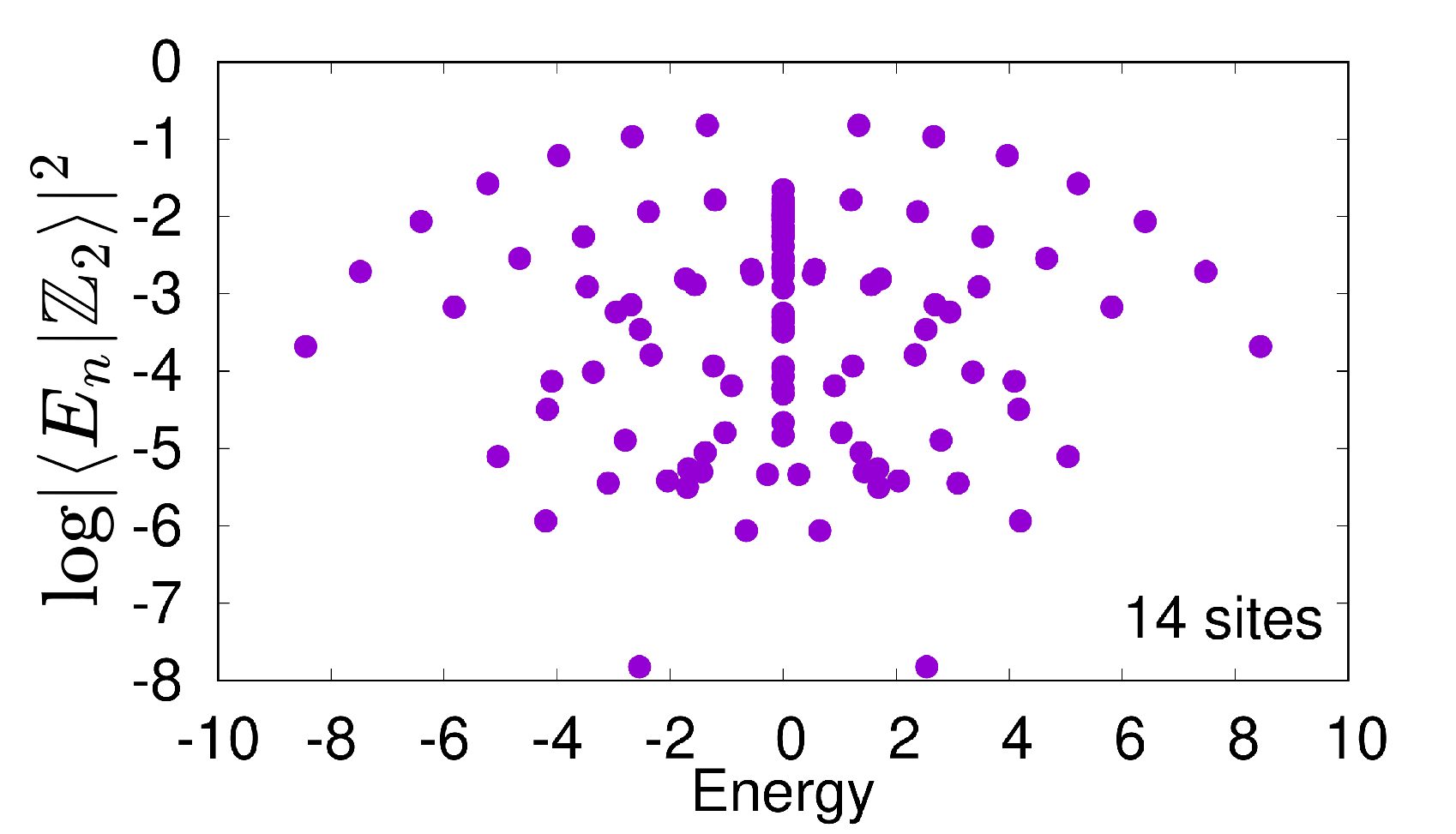}
        \caption{}
        \label{subsz2}
    \end{subfigure}%
    \hfill 
    \begin{subfigure}[b]{0.5\textwidth}
        \centering
        \includegraphics[scale=0.21]{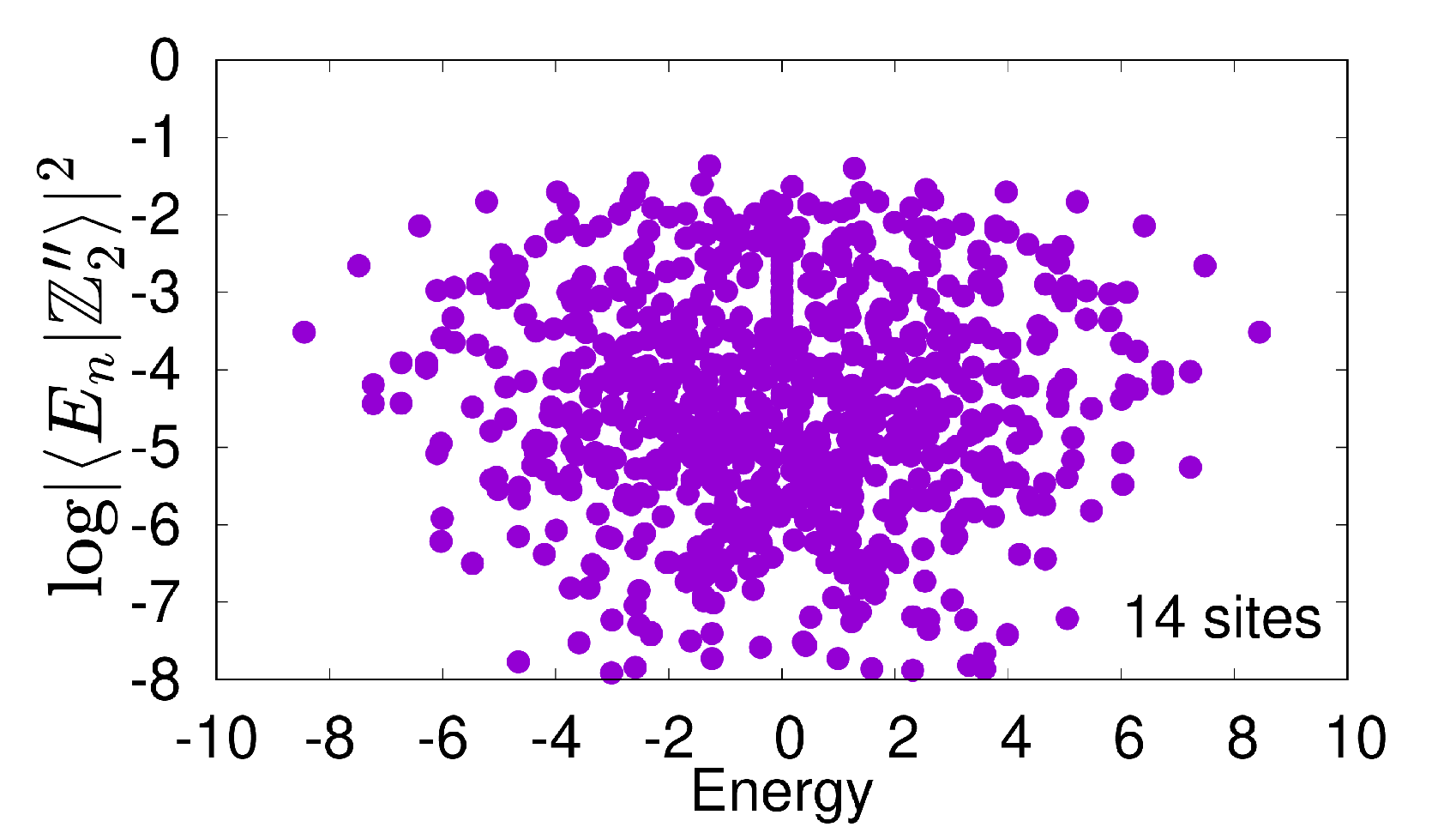}
        \caption{}
        \label{subsz22}
    \end{subfigure}
    \caption{Overlap of the eigenstates from the constrained subspace with (a) $\ket{\mathds{Z}_2}$ and  (b)$\ket{\mathds{Z}_2^{''}}$}
    \label{dilutionZ2}
\end{figure}

Conversely, in the second scenario, where we flip a spin from up to down, the resulting state remains in the constrained subspace. However, as illustrated in \Cref{dilutionZ2}, when comparing with $\ket{\mathds{Z}_2}$, which predominantly overlaps with scar states identified by their equally spaced energies, the state $\ket{\mathds{Z}_2^{''}}$ appears highly diluted across different eigenstates. Contrary to what was observed in $\ket{\mathds{Z}_2^{'}}$, ``freezing'' effect is absent, and the state can explore the constrained subspace more freely.
The combination of both lack of a coherent structure that is characteristic of scar states and local spin stability, the state is subject to the typical ergodic behavior of many-body systems, where the components of the state spread out with the full spectrum of eigenstates (in the constrained subspace), resulting in a quicker approach to thermal equilibrium.

% Conversely, in the second scenario, where we flipped a spin from up to down, the resulting state remains within the constrained subspace. Therefore, this action does not lead to a similar ``freezing'' effect. Instead, the whole chain remains dynamic, allowing for continued evolution. This lack of stabilization causes the fidelity to decrease much faster, as the system's states can change more freely and thus deviate more rapidly from the initial state. This behavior aligns with the dilution of the state within the subspace. As illustrated in  \Cref{dilutionZ2} for 14 sites, $\ket{\mathds{Z}_2}$ has an overlap predominantly with scar states, that can be identified by their equally spaced energy, which explains the known revivals of this special state. In contrast,  $\ket{\mathds{Z}_2^{''}}$  is highly diluted across different eigenstates, that can lead to a faster thermalization.

\begin{figure}[hbt!]
    \centering
    \begin{subfigure}[b]{0.5\textwidth}
        \centering
        \includegraphics[scale=0.22]{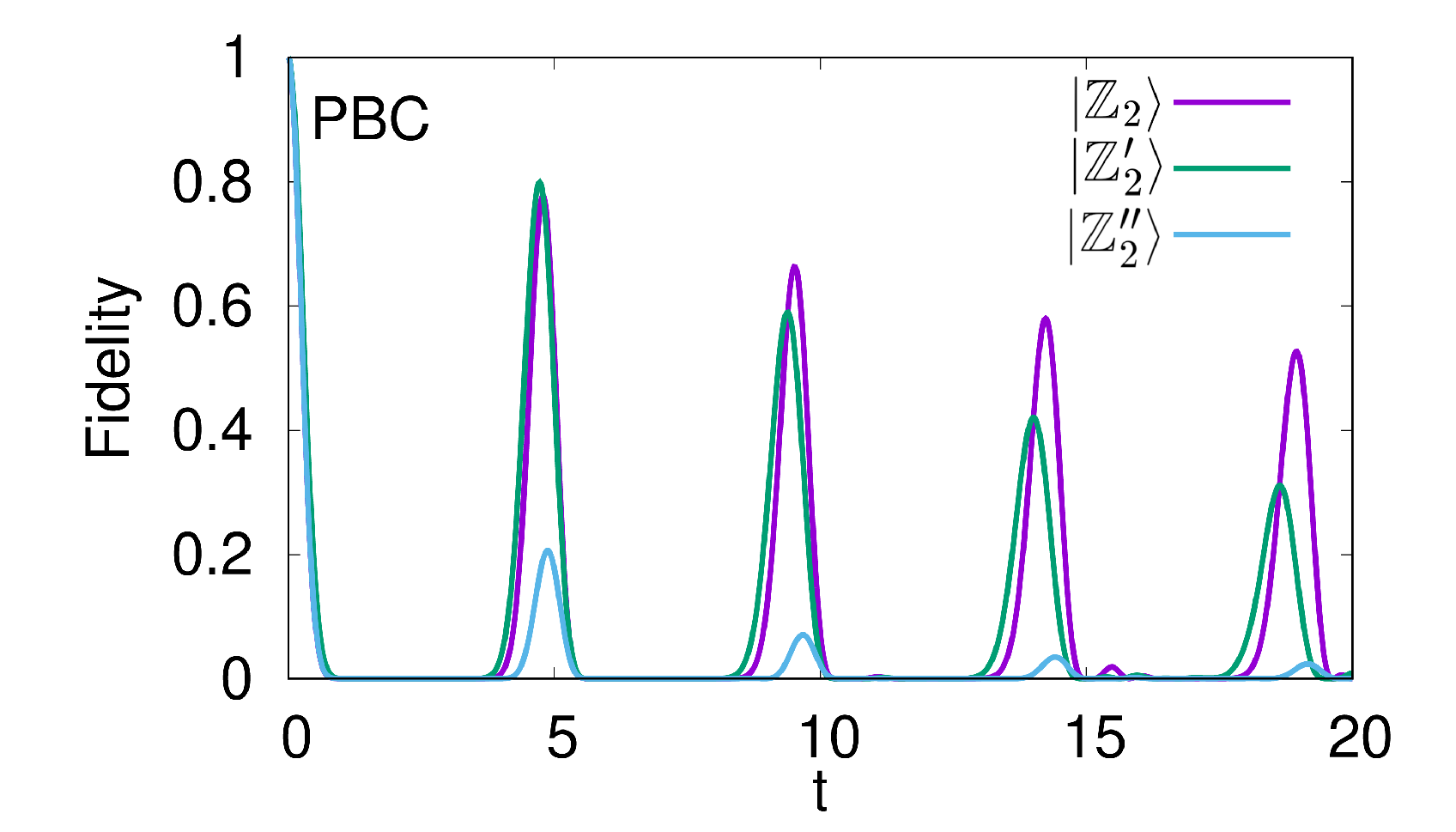}
        \caption{Fidelity }
        \label{FidelityPert}
    \end{subfigure}%
    \hfill 
    \begin{subfigure}[b]{0.5\textwidth}
        \centering
        \includegraphics[scale=0.22]{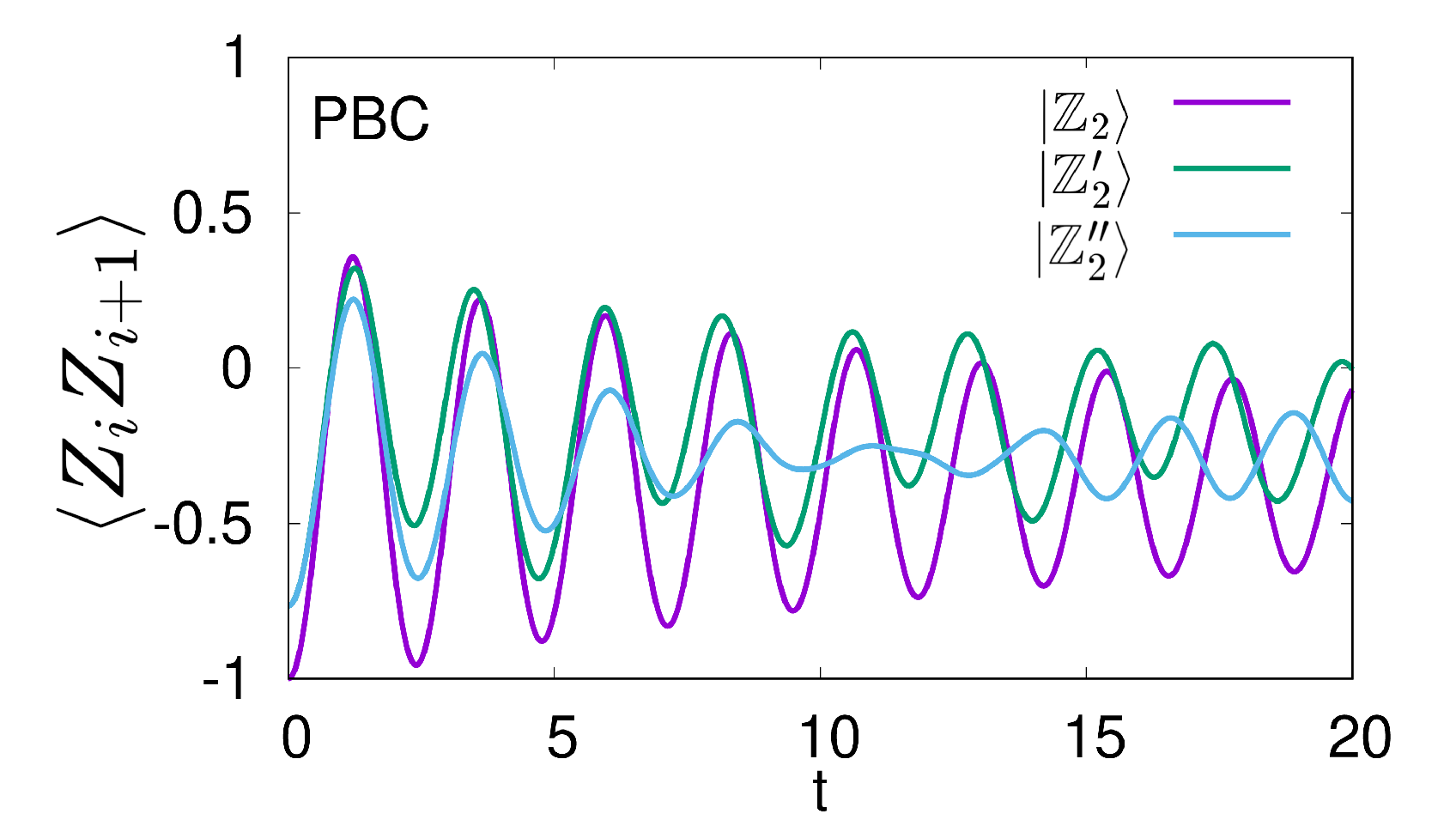}
        \caption{Correlation average }
        \label{CorrelationPert}
    \end{subfigure}
    \caption{Comparison of unperturbed $\ket{\mathds{Z}_2}$ and both possible perturbation configuration $\ket{\mathds{Z}_2^{'}}$  and $\ket{\mathds{Z}_2^{''}}$ for 18 sites }
    \label{PerturbIn}
\end{figure}

In fact, we can see in \Cref{FidelityPert} the faster decrease of the fidelity for $\ket{\mathds{Z}_2^{''}}$ in comparison to the unperturbed $\ket{\mathds{Z}_2}$, and the $\ket{\mathds{Z}_2^{'}}$with partially frozen chain keeping the occurrence of revivals. While the state $\ket{\mathds{Z}_2^{'}}$ falls outside the constrained subspace, it could intuitively suggest a faster thermalization due to a larger accessible space during dynamics and the fact that it is not in the same subspace of the scar states, however, removing the frozen spins, it can be mapped into an OBC chain of $13$ sites within the constrained subspace. As a sanity check, we simulated this case and the fidelity matched perfectly.

Regarding the average correlation, as we can see on \Cref{CorrelationPert}$,\ket{\mathds{Z}_2^{'}}$  kept more robust oscillations then $\ket{\mathds{Z}_2^{''}}$, compatible with the fidelity behavior. However, comparing the results to the $13$-site OBC mapping is challenging due to the inherent differences in the initial correlation behavior. These differences arise from the distinct number of sites and unique symmetries of each configuration. In the 18-site system, we have three frozen spins up, resulting in a difference in correlation symmetry since the start of the evolution. A second reason is that even if we did not have the frozen spins, the fact that in one case we have a chain with an even number of spins and the other is odd results in different values of the correlation. Consequently, making a direct comparison between the 13-site OBC mapping and 18 site PBC with constrained configuration, not so straightforward for this measurement.

%%%%%%%%%%%%%%%%%%%%%%%%%%%%%%%%%%%%%%%%%%%%%%%%
\section{Conclusion \label{conclusion}}

In this study, we have examined the stability and dynamics of quantum many-body scars within the PXP model, focusing particularly on their response to perturbations. We computed fidelity and average correlations to monitor the system's evolution and investigate the resilience of revivals and oscillations.

Our findings reveal that the entanglement entropy of PXP scars exhibits significant sensitivity to perturbations, approaching the profiles expected for thermal states even for very small perturbations. However, other scar signatures, such as the revivals of states having large overlap with scars, show remarkable robustness. Special states, such as $\ket{\mathds{Z}_2}$ and $\ket{\mathds{Z}_3}$, demonstrated a notable resilience to small perturbations, as evidenced by the slower dynamics observed at lower values of $W$. This suggests that these states are resistant to minor changes in the system's parameters and could potentially survive in open systems.

Furthermore, we explored the impact of minor disturbances on the $\ket{\mathds{Z}_2}$ state that previously had high overlap with scars and consistent revivals. The results indicated that different types of disturbances could lead to significantly varied behaviors. The configuration that removes the state from the constrained subspace were found to partially ``freeze'' the chain, leading to sustained oscillations and a prolonged deviation from thermalization. Additionally, this choice of configuration led to a behavior that matched a smaller chain with open boundary conditions.
Conversely, the disturbance that kept the state in the constrained subspace, showed a higher spreading into the spectrum, and accelerated the process of thermalization. The total difference of behavior highlights the sensitivity of the system to the nature of the perturbation.

These results emphasize the importance of understanding the specific conditions under which many-body scars can persist or dissipate. The resilience of certain initial states to small perturbations suggests potential applications in several areas. In quantum information processing, the ability to maintain non-thermalizing states can be leveraged to store and manipulate quantum information with reduced decoherence, enhancing the performance of quantum memories and processors~\cite{kolb2023stability}. In quantum computing ~\cite{desaules2023robust}, for example, quantum many-body scars could be used to design qubits that are more resistant to noise and perturbations, leading to more stable and reliable quantum computers. Furthermore, understanding the mechanisms that allow scars to resist thermalization can inform the creation of new quantum materials and systems that exhibit unique thermal properties, potentially useful for advanced sensing and metrology~\cite{contreras2021quantum}. Additionally, insights into the behavior of quantum many-body scars can deepen our understanding of ergodicity breaking and non-equilibrium dynamics in complex quantum systems.
Our work provides valuable insights into the mechanisms underpinning the stability of many-body scars, which could inform future research and practical applications in designing robust quantum systems.

%%%%%%%%%%%%%%%%%%%%%%%%%%%%%%%%%%%%%%%%%%%%%%

\section{Acknolegments}
This work has been supported by National Council for Scientific and Technological Development (CNPq).

\bibliography{apssamp}% Produces the bibliography via BibTeX.

\end{document}